\documentstyle[aps,prl,preprint]{revtex}
\tightenlines 

\include{psfig} 
\begin{document} \draft \title{\bf Molecular mode-coupling
theory for supercooled liquids: Application to water}

\author{L. Fabbian$^{(1)}$, A. Latz$^{(2)}$, R.
Schilling$^{(2)}$, F. Sciortino$^{(1)}$,  P. Tartaglia$^{(1)}$, C. Theis$^{(2)}$}

\address{$^{(1)}$ Dipartimento di Fisica and Istituto Nazionale per la
Fisica della Materia, Universit\'a di Roma {\it La Sapienza}, P.le Aldo
Moro 2, I-00185, Roma, Italy $^{(2)}$ Institut f\"ur Physik, Johannes
Gutenberg--Universit\"at, Staudinger Weg 7, D-55099 Mainz, Germany }

\date{\today} \maketitle

\begin{abstract} We present mode-coupling equations for the description of
the slow dynamics observed in supercooled molecular liquids close to the
glass transition. The mode-coupling theory (MCT) originally formulated to
study the slow relaxation in simple atomic liquids, and then extended to
the analysis of liquids composed by linear molecules, is here generalized to
systems of arbitrarily shaped, rigid molecules. 
We compare the predictions of the
theory for the $q$-vector dependence of the molecular
nonergodicity parameters, calculated by solving numerically the molecular
MCT equations in two different approximation schemes, with ``exact''
results calculated from a molecular dynamics simulation of supercooled
water. The agreement between theory and simulation data supports the view
that MCT succeeds in describing the dynamics of supercooled molecular
liquids, even for network forming ones.
\end{abstract}


\section{Introduction} \label{sec:introduction}

In the last years, the problem of a detailed theoretical description of the
dynamics of {\it molecular} supercooled liquids has been at the center of
renewed interest. The success of MCT
\cite{review-glass,goetze,schilling-review,yip-band,gotze-pisa,cummins-pisa}
for simple liquids in describing the weakly supercooled regime has
stimulated a considerable amount of work to extend this approach to
molecular liquids. Recent contributions include the extension of MCT to
describe the rotational dynamics of one linear probe molecule in an atomic
liquid \cite{franosch-dumb}, the extension of MCT to describe the dynamics
--- both self and collective --- of liquids of linear molecules
\cite{schilling,theis} and the MCT extension to treat the dynamics of a
full molecular systems using a site-site representation \cite{site-MCT}.
In the first part of this article we report one further step forward in the
description of the dynamics of supercooled molecular liquids, i.e. the
extension of MCT to describe the orientational and translational dynamics
for liquids composed by rigid molecules of {\em arbitrary} shape.
Within the framework of "nonlinear fluctuating hydrodynamics" this has
already been achieved by Kawasaki \cite{kawasaki97}. However, the
connection with the {\em molecular} correlation functions (see below),
which are important also from an experimental point of view has not
been worked out. This extension will be called {\em 
molecular mode coupling theory} (MMCT).

The theoretical predictions formulated in all these approaches, and the
proposed approximations are currently under investigations by several
research groups. Detailed tests of all these approaches are requested to
estimate the quality of the approximations, the predictive power of these
different approaches (i.e. the detail of the theoretical descriptions) as
well as the limit imposed by the complexity of the MCT equations and the
feasibility of their solution. Furthermore, the quality of the
approximations may depend on the molecular liquid under investigations,
for example on the fragility of the liquid in Angell's classification
scheme.

In the second part of this article we present the $q$-dependence of
molecular nonergodicity parameters calculated from the
MMCT equations for a model system of a supercooled molecular glass-forming
liquid. We then compare the theoretical predictions with equivalent
quantities calculated from extensive molecular dynamics (MD) simulations.
We choose to compare theory and MD results for liquid water. The choice of
water as molecular liquid is particularly relevant for testing the quality
of the MMCT approach, due to the presence of an extended network of
hydrogen bonds and to the peculiar local tetrahedral ordering. The
inter-molecular water-water interaction is defined by the SPC/E
potential\cite{SPCE}, which describes the molecule as a rigid planar body
and models the pair interactions as a sum of electrostatic and Lennard
Jones terms. Due to the partial charges of the atoms the molecule possesses
a dipole moment which is directed along the 2--fold rotational symmetry
axis. When referring to a body--fixed frame of reference for the molecule we
will always choose the direction of the dipole as $z$--axis and the
$x$--axis in the plane which is spanned by the molecule.

The test of MCT and MMCT can be done in two different ways. First, one can
investigate the validity of some qualitative predictions which are {\em
independent} of the system, i.e. which are the same for both theories.
These are e.g. the power law dependence of the various relaxation times on
$T-T_c$, where $T_c$ is the critical temperature for an ideal glass
transition, and the universal relationship between the corresponding
exponents and their connection with the so called exponent parameter
$\lambda$, etc. Secondly, one may calculate the {\em numerical} value of
e.g. $\lambda$ or of the $q$--dependence of the critical nonergodicity
parameters, critical amplitudes, etc. Since these quantities depend on the
system one has to use MMCT in case of molecular liquids. In this first sense
it has been shown in previous papers
\cite{self,collective,ssm,noi-vigo,mmct1} that mode-coupling theory appears
to be the correct theoretical framework for the description of the 
$\alpha$- relaxation behavior of SPC/E water. 
Indeed, the results from a
MD--simulation can be satisfactory compared to the predictions
of MCT. The comparison has been performed both for
center of mass (COM) self and collective
correlators\cite{self,collective,ssm} and $q$-independent rotational
correlators\cite{noi-vigo}. Moreover the {\em general} predictions of MCT
for the $q$-dependent molecular correlators $S_{ln,l'n'}(q,m,t)$
have been tested up to $l=2$ in Ref.\cite{mmct1,linda},
finding again a remarkable agreement, both from the qualitative and the
semi-quantitative point of view. This wide spectra analysis has confirmed
the validity of the universality of MCT predictions and has stimulated a
deeper, more quantitative understanding of the MCT approach, which we
present in this article.

The structure of the article is the following: In Sec. \ref{sec:mmct} we
present the complete set of MMCT equations for the slow dynamics in
supercooled homogeneous liquids composed by molecules of arbitrary shape.
In Sec. \ref{sec:approx} we discuss the two different approximations which
we employ to numerically solve the MMCT equations for the general molecule
case and discuss some of the numerical techniques used to solve the MMCT
equations. Finally, in Sec. \ref{sec:data} we solve the equations for the
molecular nonergodicity parameters  $F_{ln,l'n'}(q,m)$ for SPC/E water
and we compare the calculated predictions with the corresponding quantities
evaluated from the analysis of the MD trajectories.

\section{Theory} \label{sec:mmct}

\subsection{Collective correlation functions}
\label{subsec:collective}

We consider a system of $N$ identical, rigid molecules of arbitrary shape
described by the center of mass positions $\vec{x}_j(t)$ and the
orientations given by the Euler angles $\Omega_j(t)=(\phi_j(t),
\theta_j(t), \chi_j(t))$; $j=1,2,...,N$. The microscopic density
$\rho(\vec{x},\Omega,t)=\sum_j \delta(\vec{x}-\vec{x}_j(t)) \,
\delta(\Omega,\Omega_j(t))$ in the $6N$--dimensional configuration space is
expanded with respect to the complete set of functions given by the plane
waves and the Wigner matrices $D_{m n}^l(\Omega)$ \cite{G&G} to get the
tensorial density modes
\begin{equation}
\rho_{\kappa}(\vec{q},t) \equiv 
\rho_{lmn}(\vec{q},t) = i^l (2l+1)^\frac{1}{2}
\sum_{j=1}^N e^{i \vec{q}
\vec{x}_j(t)} D_{m n}^{l \ast}(\Omega_j(t))
\label{eq:density}
\end{equation}
which are the starting point of a generalized theory.
Here $l$ runs over all positive integers including zero, and $m$ as well
as $n$ takes integer values between $-l$ and $l$. To simplify the
notation we often combine $l$,$m$ and $n$ in the superindex $\kappa =
(l,m,n)$. Our aim is to give a close set of equations for the matrix ${\bf
S}$:

\begin{equation}
S_{\kappa,\kappa'}(\vec{q},t)=\frac{1}{N} \langle
\rho_{\kappa}^\ast(\vec{q},t) \rho_{\kappa'}(\vec{q}) \rangle
\end{equation}
of molecular correlation functions. The strategy to derive
such equations of motion is the same as was used for atomic liquids and
linear molecules and therefore explicit calculations shall not be given
here.  A survey of the MCT formalism for simple liquids can be found in
\cite{review-glass} and detailed calculations for a molecular system can be
found in \cite{schilling}. We restrict ourselves to pointing out where
modifications have to be made in order to treat molecules of arbitrary
shape.

Already in the choice of "slow variables" for the projection operator
formalism \cite{H&McD,forster} one is forced to make some further
considerations. In the MCT for atomic one-component liquids one uses the
density modes $\rho(q)$ and the {\em longitudinal} 
current density modes $j(q)$.
For molecular liquids this has to be modified. Besides translational
degrees of freedom (TDOF) the molecules possess also orientational degrees
of freedom (ODOF) and thus we will have to consider two different kinds of
current densities. The translational current density modes
\begin{equation}
j_{\kappa}^{T\mu}(\vec{q},t)= i^l (2l+1)^\frac{1}{2}
\sum_{j=1}^N \dot{x}_j^\mu(t) e^{i \vec{q} \vec{x}_j(t)} D_{m n}^{l
\ast}(\Omega_j(t)) \label{eq:Tcurrents}
\end{equation}
describe the
change of the density $\rho_{\kappa}(\vec{q},t)$ due to translational
motion of molecules, while the rotational current modes
\begin{equation}
j_{\kappa}^{R\mu}(\vec{q},t)= i^l (2l+1)^\frac{1}{2} \sum_{j=1}^N
{\omega'}_j^\mu(t) e^{i \vec{q} \vec{x}_j(t)} D_{m \, n+\mu}^{l
\ast}(\Omega_j(t)) \label{eq:Rcurrents}
\end{equation}
are responsible
for the change of the density due to molecular reorientation. Here
$\vec{\omega}'_j$ denotes the angular velocity in the body fixed frame. 
Consequently also $j^{R \mu}_{\kappa}$ is a vector in the body fixed
frame. We have skipped the prime because of notational reasons.
As in the theory of angular momentum in quantum mechanics it is more suitable
to use spherical components $\mu=0,\pm 1$, defined by $a^0=a_z, \, a^{\pm
1} = \frac{1}{\sqrt{2}} (a_x \pm i a_y)$, instead of the usual cartesian
vector components $a_x,a_y,a_z$. 
The connection between density and current modes
is given by the continuity equation
\begin{equation}
\dot{\rho}_{\kappa}(\vec{q},t)= \sum_{\alpha=T,R} \sum_{\mu=-1}^1
q_{ln}^{\alpha \mu \ast}(\vec{q}) \, j_{\kappa}^{\alpha \mu}(\vec{q},t)
\label{eq:continuity}
\end{equation}
with the $m$--independent coefficients
\begin{equation}
q_{ln}^{\alpha \mu}(\vec{q})= \left\{
\begin{array}{c@{\quad}c} \frac{1}{\sqrt{2}} (q_x \pm i q_y) & \alpha = T,
\mu = \pm 1 \\ q_z & \alpha = T, \mu = 0 \\ & \\
\frac{1}{\sqrt{2}}
\sqrt{l(l+1)-n(n+\mu)} & \alpha = R, \mu \pm 1 \\ n & \alpha = R, \mu = 0
\\ \end{array} \right. \label{eq:coeffs}
\end{equation}
Apart from the
separation into translational and rotational currents that was already
necessary in the case of the liquid of linear molecules we will consider
{\em all} components of the currents as slow variables instead of
the longitudinal translational currents $j_{\kappa}^T(\vec{q},t)
= 1/q \sum_{\mu} q^{T \mu \ast}_{l n}(\vec{q})
\, j_{\kappa}^{T \mu}(\vec{q},t)$ and
combined rotational currents $j_{\kappa}^R(\vec{q},t) = 1/\sqrt{l(l+1)}
\sum_\mu q_{l n}^{R \mu \ast}(\vec{q}) j_{\kappa}^{R \mu}(\vec{q},t)$ only.

This increase of the number of relevant variables for the projection
operator formalism is enforced by the following considerations.
For general molecules possessing three distinct
moments of inertia $I_1 \not= I_2 \not= I_3$ it is an important
to consider the {\em single} components $j_{\kappa}^{R \mu}$
as slow variables instead of $j^R_\kappa$ from above, because only in
the first case the long time dynamics becomes independent of the moments of
inertia, which is demanded by experimental observations.
It is also sensible to consider the components $j^{T \mu}_\kappa$ of the
translational currents separately since the evolution of the density
correlators is influenced by dynamic correlations of {\em all} components
of the currents as can be seen from the following equation:
\begin{equation}
\frac{d^2}{dt^2} \langle \rho^\ast_\kappa(\vec{q},t)
\rho_{\kappa'}(\vec{q}) \rangle = - \sum_{\alpha \alpha'} \sum_{\mu \mu'}
q^{\alpha \mu}_{l n}(\vec{q}) q^{\alpha' \mu' \ast}_{l' n'}(\vec{q})
\langle j^{\alpha \mu \ast}_\kappa(\vec{q},t) j^{\alpha'
\mu'}_{\kappa'}(\vec{q}) \rangle
\label{eq:explaination}
\end{equation}
in which terms $\langle j^{\alpha \mu \ast}_\kappa(\vec{q},t) j^{T
\pm 1}_{\kappa'}(\vec{q}) \rangle \not= 0$ occur. This dynamic coupling
to the transversal currents is induced by the anisotropy of the molecule
and exists also in the case of linear molecules.

The projection operator formalism then leads to the following continued
fraction for the Laplace transform ${\bf S}(\vec{q},z) = i \int_0^\infty
{\bf S}(\vec{q},t) e^{i z t}, \,\, Im z > 0$ of the molecular correlation
functions:
\begin{equation}
{\bf S}(\vec{q},z)=-\left[ z {\bf
S}^{-1}(\vec{q}) + {\bf S}^{-1}(\vec{q}) {\bf K}(\vec{q},z) {\bf
S}^{-1}(\vec{q}) \right]^{-1}
\label{eq:motion1}
\end{equation}
\begin{equation}
K_{\kappa,\kappa'}(\vec{q},z)=\sum_{\alpha \alpha'} \sum_{\mu \mu'}
q_{ln}^{\alpha \mu}(\vec{q}) q_{l'n'}^{\alpha' \mu' \ast}(\vec{q})
k_{\kappa,\kappa'}^{\alpha \mu, \alpha ' \mu '}(\vec{q},z)
\label{eq:motion2}
\end{equation}
\begin{equation}
\underline{{\bf k}}(\vec{q},z)=- \left[ z
\underline{{\bf J}}^{-1}(\vec{q}) + \underline{{\bf J}}^{-1}(\vec{q})
\underline{{\bf M}}(\vec{q},z) \underline{{\bf J}}^{-1}(\vec{q})
\right]^{-1}
\label{eq:motion3}
\end{equation}
where
$J_{\kappa,\kappa'}^{\alpha \mu, \alpha ' \mu '}(\vec{q},z) = 1/N \langle
j_{\kappa}^{\alpha \mu \ast}(\vec{q}) j_{\kappa'}^{\alpha' \mu'}(\vec{q})
\rangle$ is the matrix of static current correlations. The reader should
note that the under-bar of $\underline{{\bf k}}, \underline{{\bf J}}$ and
$\underline{{\bf M}}$ stands for the additional superscripts 
$\alpha \mu$ and $\alpha' \mu'$.

Thus the derivation of an equation of motion for ${\bf S}(\vec{q},t)$ has
been reduced to finding an expression for the memory kernel
$\underline{{\bf M}}(\vec{q},t)$, also called fluctuating force kernel,
since it is the correlation function of a fluctuating force. The
approximation scheme of MCT consists of a separation of $\underline{{\bf
J}}^{-1}(\vec{q}) \underline{{\bf M}}(\vec{q},t) \underline{{\bf
J}}^{-1}(\vec{q})$ into a regular part $\underline{{\bf
m}}^{reg}(\vec{q},t)$ which accounts for the fast motion and a
contribution $\underline{{\bf m}}(\vec{q},t)$ 
due to slow pairs of density modes. For an evaluation of
$\underline{{\bf m}}(\vec{q},t)$ the fluctuating force is projected onto
pairs of density modes $\rho_{\kappa_1}(\vec{q}_1)
\rho_{\kappa_2}(\vec{q}_2)$. This projection which introduces the first
approximation leads to a {\em time--dependent} four--point correlator and a
vertex function which involves {\em static} two--, three-- and four--point
correlators. In a second approximation both four--point correlators, static
and dynamic ones, are factorized into a product of two--point
correlators. In a final step one can approximate the static three--point
correlator by use of the generalized convolution approximation which is
easily generalized from linear \cite{schilling} to arbitrary molecules.
As already pointed out above we will not
give the complete derivation here since it is rather involved and
analogous to the case of linear molecules. The result for $\underline{{\bf
m}}(\vec{q},t)$ can be expressed as follows:
\begin{eqnarray}
m^{\alpha \mu, \alpha' \mu'}_{\kappa, \kappa'}(\vec{q},t) &=&
\frac{1}{2N} \left(\frac{\rho_0}{8 \pi^2} \right)^2
{\sum_{\vec{q}_1,\vec{q}_2}}' \sum_{\kappa_1,\kappa'_1}
\sum_{\kappa_2,\kappa'_2}
v^{\alpha \mu}_{\kappa \kappa_1 \kappa_2}
(\vec{q},\vec{q}_1,\vec{q}_2) \,
v^{\alpha' \mu' \ast}_{\kappa' \kappa'_1
\kappa'_2}(\vec{q},\vec{q}_1,\vec{q}_2) \; \times \nonumber \\
& & \; \times \; 
S_{\kappa_1, \kappa'_1}(\vec{q}_1,t) 
S_{\kappa_2, \kappa'_2}(\vec{q}_2,t).
\label{eq:memory1}
\end{eqnarray}
Here the prime denotes summation such that $\vec{q}_1+\vec{q}_2=\vec{q}$.
Besides the time--dependent molecular correlators there appear the number
density $\rho_0 = N/V$ and the vertex functions which are products of
\begin{equation}
v^{\alpha \mu}_{\kappa \kappa_1
\kappa_2}(\vec{q},\vec{q}_1,\vec{q}_2) = \sum_{\kappa_3} u^{\alpha
\mu}_{\kappa \kappa_3 \kappa_2}(\vec{q},\vec{q}_1,\vec{q}_2) 
c_{\kappa_3, \kappa_1}(\vec{q}_1) + (1 \leftrightarrow 2)
\label{eq:memory2}
\end{equation}
where ${\bf c}(\vec{q})$ denotes the matrix of direct
correlation functions which is related to the static structure factors by
\begin{equation}
{\bf c}(\vec{q}) = \frac{8 \pi^2}{\rho_0} ({\bf 1} - {\bf
S}^{-1}(\vec{q})).
\label{eq:directcf}
\end{equation}
The coefficients $u$ are given by
\begin{eqnarray}
u^{\alpha \mu}_{\kappa \kappa_1 \kappa_2}(\vec{q},\vec{q}_1,\vec{q}_2)
&=& i^{l_1+l_2-l} \left( \frac{(2l_1+1) (2l_2+1)}{2l+1} \right)^\frac{1}{2} 
{\cal C}(l_1 l_2 l; m_1 m_2 m)
\quad \times \nonumber \\ & & \; \times \;
q_{l_1 n_1}^{\alpha \mu \ast}(\vec{q}_1) \left\{
\begin{array}{c@{\quad}c}
{\cal C}(l_1 l_2 l; n_1 n_2 n) & \alpha = T \\
{\cal C}(l_1 l_2 l; n_1+\mu \, n_2 \, n+\mu) & \alpha = R
\end{array} \right.
\label{eq:memory3}
\end{eqnarray}
and ${\cal C}(l_1 l_2 l; m_1 m_2 m)$ denotes the usual Clebsch--Gordan
coefficients \cite{G&G}.

The equations (\ref{eq:motion1}-\ref{eq:motion3}) together with equation
(\ref{eq:memory1}) form a closed set of equations 
for the correlation matrix ${\bf
S}(\vec{q},t)$. The vertex functions given by equations
(\ref{eq:memory2}-\ref{eq:memory3}) are determined by the number
density $\rho_0$ and the static structure factors ${\bf S}(\vec{q})$, only.

\subsection{Tagged particle correlation function}
\label{subsec:self}

We will now examine the motion of a {\em single} 
molecule that is immersed in a
molecular liquid. Again we have a liquid of $N-1$ identical, rigid
molecules of mass $M$ and moments of inertia $I_1$,$I_2$,$I_3$ whose center
of mass coordinates are denoted as $x_j(t)$ and whose orientations are
given by the Euler angles $\Omega_j(t)$. In addition we have, as the $N$-th
particle, a molecule of mass $M'$ and moments of inertia $I'_1$,$I'_2$,$I'_3$.
As a special case we get the self correlator of a selected
particle within a homogeneous liquid if the tagged particle is identical to
the molecules of the surrounding liquid.

Besides the quantities we have already introduced in the previous
subsection we have to consider the density modes for the tagged particle
\begin{equation}
\rho^{(s)}_{\kappa}(\vec{q},t) = e^{i \vec{q} \vec{x}_N(t)} D^{l \ast}_{m
n}(\Omega_N(t))
\label{eq:selfdensity}
\end{equation}
and the corresponding tagged particle correlation function
\begin{equation}
S_{\kappa,\kappa'}^{(s)}(\vec{q},t) = \langle \rho_{\kappa}^{(s)
\ast}(\vec{q},t) \rho_{\kappa'}(\vec{q}) \rangle.
\label{eq:selfcorrelator}
\end{equation}
The slow variables for the projection operator formalism
are given by the density modes (\ref{eq:selfdensity}) and the current
densities
\begin{eqnarray}
j^{(s) T \mu}_{\kappa}(\vec{q},t) &=&
\dot{x}_N^{\mu}(t) e^{i \vec{q} \vec{x}_N(t)} D^{l \ast}_{m n}(\Omega_N(t))
\label{eq:selfTcurrent} \\
j^{(s) R \mu}_{\kappa}(\vec{q},t) &=&
{\omega'}_N^{\mu}(t) e^{i \vec{q} \vec{x}_N(t)} D^{l \ast}_{m \,
n+\mu}(\Omega_N(t)) \label{eq:selfRcurrent}
\end{eqnarray}
of the tagged
particle. Also in this case it is necessary to use all components of the
rotational currents separately to avoid inertia effects in the long--time
behavior. The results of the projection operator formalism are analogous
to the equations for the coherent correlations. As a further simplification
the static self--correlations are given by
$S_{\kappa,\kappa'}^{(s)}(\vec{q})=\delta_{\kappa,\kappa'}$.
Thus we get
\begin{equation}
{\bf S}^{(s)}(\vec{q},z) = - \left[ z {\bf 1} + {\bf K}^{(s)}(\vec{q},z)
\right]^{-1} \label{eq:self1}
\end{equation}
\begin{equation}
K^{(s)}_{\kappa,\kappa'}(\vec{q},z) =
\sum_{\alpha \alpha'} \sum_{\mu \mu'} q^{\alpha \mu}_{ln}(\vec{q})
q^{\alpha' \mu' \ast}_{l'n'}(\vec{q}) k^{(s) \alpha \mu, \alpha'
\mu'}_{\kappa,\kappa'}(\vec{q},z) \label{eq:self2}
\end{equation}
\begin{equation}
\underline{{\bf k}}^{(s)}(\vec{q},z) = - \left[ z \underline{{\bf J}}^{(s)^{-1}}(\vec{q}) +
\underline{{\bf J}}^{(s)^{-1}}(\vec{q}) \underline{{\bf M}}^{(s)}(\vec{q},z)
\underline{{\bf J}}^{(s)^{-1}}(\vec{q}) \right]^{-1} \label{eq:self3}
\end{equation}
with the same coefficients $q^{\alpha \mu}_{ln}(\vec{q})$ as above.

The correlations of the tagged particle will be controlled by the coherent
correlations. Therefore in the mode coupling approximation for the memory
function $\underline{{\bf m}}^{(s)}(\vec{q},t)$, the slow part of
$\underline{{\bf J}}^{(s)^{-1}}(\vec{q}) 
\underline{{\bf M}}^{(s)}(\vec{q},t) 
\underline{{\bf J}}^{(s)^{-1}}(\vec{q})$, the fluctuating forces are
projected onto bilinear products of tagged particle and coherent density
modes. In the thermodynamic limit the effect of the tagged particle on the
surrounding liquid can be neglected and the coherent correlator is
identical to the correlation function for the homogeneous liquid. The mode
coupling approximation for the 4--point correlation functions thus leads to
the following expression for the memory function
\begin{eqnarray}
& &m^{(s) \alpha \mu, \alpha' \mu'}_{\kappa, \kappa'}(\vec{q},t)
= \left( \frac{\rho_0}{8 \pi^2} \right)^2 \frac{1}{N} 
{\sum_{\vec{q}_1,\vec{q}_2}}'
\sum_{\kappa_1,\kappa'_1} \sum_{\kappa_2,\kappa'_2}
v^{(s) \alpha \mu}_{\kappa \kappa_1
\kappa_2}(\vec{q},\vec{q}_1,\vec{q}_2) v^{(s) \alpha' \mu' \ast}_{\kappa'
\kappa'_1 \kappa'_2}(\vec{q},\vec{q}_1,\vec{q}_2)
\; \times \nonumber \\ & & \quad \times \;
S^{(s)}_{\kappa_1,\kappa'_1}(\vec{q}_1,t) 
S_{\kappa_2, \kappa'_2}(\vec{q}_2,t).
\label{eq:self4}
\end{eqnarray}
with the vertex functions
\begin{equation}
v^{(s) \alpha \mu}_{\kappa
\kappa_1 \kappa_2}(\vec{q},\vec{q}_1,\vec{q}_2) = \sum_{\kappa_3} u^{\alpha
\mu}_{\kappa \kappa_3 \kappa_1}(\vec{q},\vec{q}_2,\vec{q}_1)
c^{(s)}_{\kappa_3, \kappa_2}(\vec{q}_2).
\label{eq:self5}
\end{equation}
The coefficients $u$ are the same as given above and the direct correlation
function that describes the interaction between the tagged particle and the
surrounding liquid is defined by
\begin{equation}
\rho_0 ({\bf c}^{(s)}(\vec{q}) {\bf S}(\vec{q}))_{\kappa, \kappa'} =
\langle \rho^{(s) \ast}_\kappa(\vec{q}) \rho_{\kappa'}(\vec{q}) 
\rangle - \delta_{\kappa, \kappa'}.
\label{eq:cself}
\end{equation}
In the special case that the
tagged particle has the same properties as the molecules of the liquid
${\bf c}^{(s)}(\vec{q})$ is just the ordinary direct correlation function
of the homogeneous liquid.

Detailed investigations for the tagged particle correlators have
been done for a dumbbell molecule in a simple isotropic 
liquid \cite{franosch-dumb}.
The equations given here are the generalization of this theory (for a
linear molecule in a simple liquid) to the general case of an arbitrary
shaped molecule in a molecular liquid.

\section{Approximations} \label{sec:approx}

The aim of our numerical investigations was to examine the long--time
behavior of the solutions of the equations of motion presented in section
\ref{subsec:collective}, i.e. to calculate the critical nonergodicity
parameters ${\bf F}(\vec{q}) = \lim_{t \to\infty} {\bf S}(\vec{q},t)$ and
the transition temperature $T_c$. As input for these calculations we have
determined the static structure factors ${\bf S}(\vec{q})$ from a
MD--simulation, as described in detail in a previous publication
\cite{mmct1}. As discussed there, the $C_{2v}$--symmetry of the water
molecule leads to the consequence that the distinct part ${\bf
S}^{(d)}(\vec{q})$ of the structure factors vanishes for $n$ and/or $n'$
odd, i.e. $S_{lmn,l'm'n'}(\vec{q})$ with $n$ and $n'$ odd only contains
information about the self--correlation of the molecules. These symmetry
relations allow for a simplification of the equations. As shown in appendix
\ref{app:neven} the matrix equation splits into two parts. The matrix
elements $S_{lmn,l'm'n'}(\vec{q},t)$ with $n$ and $n'$ even are determined
by a closed set of equations which is independent of the correlators with
$n$ and/or $n'$ odd. The remaining part of the equations, i.e. for $n$ and
$n'$ odd, is identical to the corresponding tagged particle equations for
the self--correlators as presented in section \ref{subsec:self}.

For these numerical studies it is further useful to transform the equations
to the $q$-frame representation, i.e. to choose the $z$-axis of the
laboratory frame in direction of the vector $\vec{q}$. The resulting set of
equations and some details of their derivation are given in appendix
\ref{app:q-frame}. The $q$--frame offers the advantages that the
correlation matrices depend only on the modulus $q=|\vec{q}|$ and in
addition are diagonal with respect to the indices $m$ and $m'$. Thus we
have to solve self-consistently a set of equations for the nonergodicity
parameters $F_{ln,l'n'}(q,m) \equiv F_{lmn,l'mn'}(q \hat{e}_z)$.

The main computational problem in solving the equations for the glass form
factor is the calculation of the memory matrix
$m_{\kappa,\kappa'}^{\alpha \mu, \alpha' \mu'}(q)$, 
due to the enormous number of
terms in the summation of eq.(\ref{eq:memory1}).
Of course, any attempt of numerical calculation requires the introduction
of an upper cut-off $l_{co}$ in $l,l'$, in order to have a finite number of
coupled equations. 
The stable solution ${\bf F}(q,m)$ is 
found as the fixed point of the iterative
solution of the given equations.

It has been estimated in several MCT calculations that, in order to have a
reasonable convergence towards the fixed point, it is necessary to perform
several hundreds of iterations. Although it is possible to considerably
reduce the number of elements in the sum of eq.(\ref{eq:memory1}) by taking
into account the symmetries of the molecule, a full solution for $l_{co}=2$
is still not feasible.

As discussed in section \ref{sec:mmct} it is necessary to take into account
the components of the rotational currents $j^{R \mu}$ separately to avoid
inertia effects in the long--time behavior. Thus we have taken into
account all corresponding memory functions. With respect to the
translational currents we have decided to take into account only the
longitudinal components $j^{T 0}$, i.e. all memory functions with $\alpha=T,
\mu=\pm 1$ or $\alpha'=T, \mu'=\pm 1$ are neglected.

The structure of the MMCT
equations further offers the possibility of several approximation schemes,
differing in the choice of the molecular static structure factors which
are taken into account. In this article we present MMCT calculations for
two different approximations, which bring the numerical calculations to the
frontier of the present computer facilities. In both approximations we
neglect the third angular index $n$ in the static quantities, i.e. we
include as input of the calculation only the static structure factors
$S_{l0,l'0}(q,m)$ and, thus, the direct correlation functions
$c_{l0,l'0}(q,m)$. In addition, we put $F_{ln,l'n'}(q,m)$ to zero for $n$
and $n'$ different from zero. This approximation reflects in a reduction of 
the number of independent memory kernels to be calculated, although it does not
imply the simplification $n=n'=0$ in the memory functions.

Examining the intensity and the temperature dependence of the static
correlation functions we have given some justification for this
approximation in a previous publication \cite{mmct1}. Still it has to be
noted that it is mainly motivated by the need to reduce the
computational burden and we plan to put a significant effort in the
direction of a full solution of the MMCT equations, including also the
angular index $n$.

In Ref.\cite{mmct1}, it has been shown that the distinct part of the
structure factors vanishes for odd $n$ and/or $n'$. Thus, with $l \leq 2$,
the approximation $n=n'=0$  essentially means neglecting the coupling
with the correlators with $|n|=2$ and/or $|n'|=2$. This approximation is
equivalent to neglecting the third Euler angle $\chi$, i.e. the rotations
of the water molecules around the dipolar axes. This means that the motion
of the water molecules is reduced to the motion of their dipole
moments. Thus, we will refer to this first simplification as the {\it
dipole approximation}. 

An even stronger approximation is defined by assuming in addition that both
the static structure factors, the critical nonergodicity parameters and
the memory functions are diagonal in $l$ and $l'$, i.e.
\begin{equation}
\label{eq:diag_sq}
S_{l0,l'0}(q,m)=S_{l0,l0}(q,m)\delta_{ll'}
\end{equation}
\begin{equation}
\label{eq:diag_fq}
F_{l0,l'0}(q,m)=F_{l0,l0}(q,m)\delta_{ll'}
\end{equation}
\begin{equation}
\label{eq:diag_mq}
m_{lmn,l'm'n'}^{\alpha \mu, \alpha' \mu'}(q) =
m_{lmn,lm'n'}^{\alpha \mu, \alpha' \mu'}(q) \delta_{ll'}
\end{equation}
Thus ${\bf S}(q,m)$ becomes a diagonal matrix with $6$ non
vanishing elements (the diagonal ones) while 
$\underline{{\bf m}}(q)$ is still non
diagonal with respect to $\alpha \mu$ and $\alpha' \mu'$.
This approximation and also the even
stronger restriction of additional diagonality of the memory kernel with
respect to $\alpha \mu$ has also been used in the study of a dumbbell in a
simple liquid \cite{franosch-dumb} and a liquid of diatomic molecules
\cite{schilling-vigo}.

We have iteratively solved the equations for the nonergodicity parameters
on a grid of $100$ $q$-vectors ranging up to $110 nm^{-1}$. Within the
diagonal approximation, one iteration step requires about $15$ minutes on
one $\alpha$-station. The dipole-approximation requires about two hours of
CPU time and we estimated that a full solution of the MMCT equations
including $n$ would require about 4 days per iteration.

\section{results} \label{sec:data}

We have found $T_c^{MMCT}=206~K$ in the diagonal-dipole approximation. At
this temperature the solution of the nonergodicity parameter equations has
been iterated until the average difference over the whole $q$ range between
the $(n+1)$--th iteration ${\bf F}^{(n+1)}(q,m)$ and the $n$--th ${\bf
F}^{(n)}(q,m)$ was of order $10^{-10}$.  Since the MCT--approach to e.g. a
hard sphere system \cite{hardspheres}, binary liquids \cite{kob-LJ} and
diatomic molecules \cite{winkler} has shown that MCT overestimates the
freezing into a glassy state, we
consider fortuitous the agreement between the estimated $T_c^{MMCT}$
and the numerical estimate of the critical temperature from the analysis of
the molecular dynamics data, $T_c^{MD}=200 \pm 3~K$, for SPC/E water.
We want to highlight that despite the
diagonalization approximation, differently from what has been found for the
liquid of Lennard-Jones dumb-bells described in Ref.\cite{schilling-vigo},
in the case of water the theory does not yield an unphysical splitting of
rotational and translational transition temperatures, i.e. all the elements
of the theoretical nonergodicity matrix ${\bf F}(q,m)$ simultaneously jump
from zero to a nonzero value at the same temperature, or in other words:
{\em all} degrees of freedom freeze at a {\em single} temperature.

In the dipole approximation, i.e. relaxing the diagonality approximation,
the equations for ${\bf F}(q,m)$ have been solved in a similar way.
In this approximation the value of the {\em critical} ${\bf
F}(q,m)$ was evaluated stopping the iterative calculation when the average
difference between two consecutive iterations was of order $10^{-8}$. In
the dipole approximation, the theoretical critical temperature has been
found to be about ${T'}_c^{MMCT}=208~K$, which is not so different from
$T_c^{MMCT}$ in the diagonal approximation and from the numerical one. As
discussed in Ref.\cite{mmct1}, it is reasonable to suppose that the
transition is driven by the diagonal structure factors, especially the ones
with $l=0$ or $l=2$, which are the most sensitive to variations of
temperature. It is thus not surprising to observe that the critical
temperature is almost insensitive to the introduction of the off-diagonal
terms, which display a weaker dependence on $T$.

The theoretical predictions for ${\bf F}(q,m)$, in the different
approximation schemes, are shown in Figs. \ref{fig:fqZZZZZZ},
\ref{fig:fig1mct}, \ref{fig:fig2mct} and \ref{fig:figcrossmct} in
comparison  to the corresponding quantities as evaluated from the MD
simulation fitting the time evolution of the correlators in the early
$\alpha$-region to the von Schweidler law\cite{mmct1}
\begin{eqnarray}
& & S_{ln,l'n'}(q,m,t) - F_{ln,l'n'}(q,m) \cong -H^{(1)}_{ln,l'n'}(q,m)
\hat{t}^b + H^{(2)}_{ln,l'n'}(q,m) \hat{t}^{2b} + O(\hat{t}^{3b})
\label{eq:vs}
\end{eqnarray}
In Fig. \ref{fig:fqZZZZZZ} the
theoretical predictions for the COM nonergodicity parameter in the two
approximations examined in this article are compared to the MD data. Both
approximations are excellent in the low $q$ range (up to around
$30~nm^{-1}$). The MMCT approximations allow the calculation, beside the
COM glass form factor, also of the angular nonergodicity parameters. Of
course, in the diagonal dipole approximation only the diagonal elements
$F_{l0,l0}$ of the nonergodicity matrix ${\bf F}$ can be evaluated. The
results are shown in Figs. \ref{fig:fig1mct} and \ref{fig:fig2mct} together
with the predictions for the same quantities as evaluated relaxing the
diagonality approximation. The agreement with the numerical data (symbols
in the figures) is satisfactory for the $l=1$ nonergodicity parameters
(both for $m=0$ and $m=1$), while the predictions corresponding to $l=2$
are less satisfactory. It is reasonable to expect that the worse results obtained for $l=2$ are 
due to the fact that this is exactly the cut-off value, and it is thus more
sensitive to ``boundary'' effects. The comparison between theoretical
predictions and MD data for the off-diagonal terms of ${\bf F}$, a
comparison which is possible to perform only in the dipole approximation
scheme, is shown in Fig. \ref{fig:figcrossmct}. Again, the agreement
between MD data and theoretical results become worse on increasing $l$
towards $l_{co}$. By comparing the results for the two different
approximations, we note that both, for the COM (Fig. \ref{fig:fqZZZZZZ})
and angular(Figs. \ref{fig:fig1mct} and \ref{fig:fig2mct}) nonergodicity
parameters, the coupling to the  non diagonal correlators introduced in the
pure dipole approximation contribute very little to the determination of
the diagonal terms of the glass form factor. This result, as well as the
small variation of the critical temperature within the two approximations,
is due to the small amplitudes of the off-diagonal terms with respect to the diagonal ones and
supports the idea that the critical behavior of the system is mainly driven
by the more intense structure factors. This consideration suggests
to use only the strongest $S_{ln,l'n'}(q,m)$ as input of the
calculation, which may allow to choose higher values for the cut-off
$l_{co}$.

\section{Summary and Conclusions} \label{sec:conclusions}

In the present paper we have performed a quantitative test of MMCT for the
SPC/E--model for water in the supercooled regime. MMCT is an extension of
mode coupling--theory for simple liquids to molecular systems. It provides
equations of motion for the molecular correlators
$S_{\kappa,\kappa'}(\vec{q},t)$ which form a {\em complete} set for any
time--dependent two--point correlator. Of course, there are infinitely many
of them which demands for a truncation of the set of MMCT--equations at a
cut--off value $l_{co}$ for $l$ and $l'$. In our case we have chosen
$l_{co}=2$. From a pragmatic point of view this may be justified by the
fact that up to today there seems to be no experimental method which allows
to determine those correlators for $l$ and $l'$ larger than two, although
this can be done for any numerical simulation. Instead of using the
molecular correlators one also could use correlators in a {\em site--site}
representation \cite{H&McD}. MCT for molecular systems in a site--site
description has recently been worked out \cite{site-MCT}. This type of
approach has the advantage that for molecules with a few atoms the number of
correlators is small, e.g. for water there are six correlators in maximum,
where one has to take into account that both protons can be distinguished
for a classical system. A site--site description, however, has the
disadvantage that it is incomplete, i.e. information has been lost. For
instance light scattering, dielectric spectroscopy, NMR, ESR, etc. directly
yield {\em molecular} correlators. Whereas the
site--site correlators can be represented by the molecular ones, the
reverse is not possible \cite{neutron}. Hence, from a site--site
description no {\em exact} information can be deduced for the experimental
techniques mentioned above. Nevertheless, it would be worthwhile to solve
the molecular MCT--equations in the site--site representation, e.g. for
water, in order to compare the critical temperature $T_c$, the critical
nonergodicity parameters, etc. with the corresponding quantities from MMCT.

Despite the cut--off for $l$ and $l'$ the MMCT--equations are still rather
involved.
Therefore we decided to perform in a first step a further
approximation which is that $n$ and $n'$ is put to zero for the static
correlators and the nonergodicity parameters. Because
of this approximation we neglect rotations of both protons around the
molecular symmetry axis. Within these approximations we have calculated the
(unnormalized) critical nonergodicity parameters $F_{l0,l'0}(q,m)$ in the
$q$--frame by solving the corresponding MMCT--equations. As we have found
that the {\em diagonal}, {\em static} correlators $S_{l0,l'0}(q,m)$,
compared with the nondiagonal ones, are most dominant with respect to their
intensity and temperature--dependence \cite{mmct1}, we have additionally
solved the MMCT--equations by assuming all static correlators and the
nonergodicity parameters to be diagonal in $l$ and $l'$. The reader should
note that the diagonality of the nonergodicity parameters is an additional
approximation. This latter approximation has been motivated by a similar
investigation for a supercooled liquid of diatomic molecules
\cite{schilling-vigo}.

The solution of the MMCT--equations yield the critical temperature
${T'}_c^{MMCT} \cong 208 K$. In the diagonalization approximation we also
obtained a {\em single} transition temperature $T_c^{MMCT} \cong 206 K$,
which does not differ much from ${T'}_c^{MMCT}$ and $T_c^{MD} \cong 200 \pm 3
K$. We consider this very good agreement as fortuitous since usually the
mode coupling theory strongly overestimates the freezing
\cite{kob-LJ,winkler}. That the diagonalization approximation almost does
not affect the transition temperature is quite different to what has been
found from a similar study for diatomic molecules with Lennard--Jones
interactions. There the diagonalization approximation causes the separate
freezing of the COM--correlators ($l=l'=0$) and the "orientational"
correlators with $l=l'\not=0$ \cite{schilling-vigo}. This quite different
behavior is probably related to the much stronger
translational--orientational coupling in case of water.

Comparison of the MMCT--results for $F_{l0,l'0}(q,m)$ with the
corresponding MD--results leads to the following main conclusions:
\begin{enumerate}
\item $F_{l0,l'0}(q,m)$ obtained from MMCT, without and with
diagonalization approximation, differ only slightly from each other which
confirms the dominance of the diagonal correlators.
\item The qualitative $q$--dependence of $F_{l0,l'0}(q,m)$ from the
MD--simulation is well reproduced by the corresponding MMCT--result. This
is even true for some fine structure like the shoulder at $q \cong 20
nm^{-1}$ of $F_{10,10}(q,m=0)$.
\item A good {\em quantitative} agreement between the MD-- and
MMCT--results is found for the full $q$--range for $F_{10,10}(q,m=1)$ and
$F_{10,00}(q,m=0)$. For the other cases, except those with $l=l'=2$, a
reasonable agreement is found for $q < 30 nm^{-1}$. For $F_{20,20}(q,m)$
the deviations are rather large, particularly for $q < 40 nm^{-1}$.
\end{enumerate}
This behavior for $F_{l0,l'0}(q,m)$ is in full accordance with that for
diatomic molecules \cite{schilling-vigo,winkler}. 
The larger discrepancy for the case
$l=l'=2$ is probably due to the cut--off at $l_{co}=2$. For a single
dumbbell in an isotropic liquid of hard spheres it has been shown that MMCT
yields accurate results for e.g. $F_{l0,l0}(q,m)$ if one chooses
$l_{co}=l+2$ \cite{goetzepc}.

Since this first quantitative comparison of MMCT and the MD--results for
water is encouraging we are planning to extend our MMCT--study for water to
include $n$ and $n'$. This will offer the possibility to answer the
interesting question whether the  $180^\circ$--jumps of the protons,
which leave the molecule invariant, freeze at the same critical temperature
where all the other degrees of freedom freeze, or if they will freeze at a
lower temperature. Besides the nonergodicity parameters it would also be
interesting to calculate from MMCT the critical amplitudes
$H_{ln,l'n'}(q,m)$ and the exponent parameter $\lambda$ from which the
critical exponent $a$, the von Schweidler exponent $b$ as well as $\gamma$
which characterizes the power law divergence of the $\alpha$--relaxation
timescale, can be obtained. Finally a solution of the time--dependent
MMCT--equations would be desirable. This extensions will be hard to achieve
without further approximation schemes. Our results give a strong indication
that this could be possible. As the comparison of the different
approximations presented in this paper demonstrates, the solutions of the
MMCT are not strongly affected, if small and temperature insensitive
components of the {\em static} structure factors $S_{ln,l'n'}(q,m)$ are
neglected. Therefore it should be possible to restrict the MMCT to the most
relevant components, where the question of relevance is decided on the
basis of the static correlation functions. Depending on the system, this
procedure can lead to a dramatic reduction of memory functions to be
calculated. This could enable us to capture more qualitative features of
the nonergodicity parameters and dynamics, by including the relevant
components with $l > 2$, thereby neglecting irrelevant components with $l
\le 2$.

To summarize, we can say that this first quantitative test of MMCT for water
has demonstrated reasonable agreement of both, the critical temperature and
critical nonergodicity parameters obtained from MMCT and a MD--simulation,
although the agreement of the transition temperatures should not be
overestimated.

%
%
%
%
%
%
%
%
%

\section{acknowledgment}

Research of L.F., F.S. and P.T is supported by MURST (prin 98). A.L., R.S.
and C.T. gratefully acknowledge financial support by SFB 262.

\begin{appendix}

\section{Simplification of the MMCT--equations for water molecules}
\label{app:neven}

The $C_{2v}$--symmetry of the water molecule leads to the consequence that
the molecular correlation function 
$S_{\kappa,\kappa'}(\vec{q},t)$ for $n$ and
$n'$ odd are given by the self--correlations and vanish for $n$ odd, 
$n'$ even or $n$ even,$n'$ odd,
i.e. the matrix ${\bf S}(\vec{q},t)$ has the block structure
\begin{equation}
\begin{array}{c|c|c|c}
 & n' \; \mbox{even} & n' \; \mbox{odd} & \\ \hline 
 & \tilde{{\bf S}}(\vec{q},t) & {\bf 0} & n \; \mbox{even} \\
{\bf S}(\vec{q},t) = & & & \\
 & {\bf 0} & \tilde{{\bf S}}^{(s)}(\vec{q},t) & n \; \mbox{odd}
\end{array}
\label{eq:Sblocks}
\end{equation}
Thus the direct correlation functions $c_{\kappa,\kappa'}(\vec{q})$
are nonzero for $n$ and $n'$ even, only. From equation (\ref{eq:memory2}) we
can conclude that the functions $v^{\alpha \mu}_{\kappa \kappa_1
\kappa_2}(\vec{q},\vec{q}_1,\vec{q}_2)$ vanish if $n_1$ and $n_2$ are odd.
As a consequence the memory functions can {\em not} contain terms of the kind
$v * v * \tilde{S}^{(s)} * \tilde{S}^{(s)}$.
Using that the factors $u$ (see eqs. (\ref{eq:memory2}),
(\ref{eq:memory3}))
contain the Clebsch--Gordan coefficients ${\cal C}(l_1 l_3 l; n_1 n_3 n)$ or
${\cal C}(l_1 l_3 l; n_1+\mu \, n_3 \, n+\mu)$ or the corresponding ones with
$(1 \leftrightarrow 2)$, where $n_3$ has to be even since it occurs also as
index of the direct correlation function (see eq.(\ref{eq:memory2})) 
one concludes that terms of the
kind
\begin{equation}
v \, * \, v \, * \, \tilde{S} \, * \, \tilde{S}
\label{eq:type1}
\end{equation}
are only contained in memory functions
with $n$ and $n'$ even. Further one finds, that memory functions with $n$
and $n'$ odd contain only terms of the kind
\begin{equation}
v \, * \,  v \, * \, \tilde{S} \, * \, \tilde{S}^{(s)}
\label{eq:type2}
\end{equation}
while memory functions with different $n$ even,$n'$ odd or $n$ odd, 
$n'$ even vanish.
We can summarize those findings in the schematic representation
\begin{equation}
\begin{array}{c|c|c|c}
 & n' \; \mbox{even} & n' \; \mbox{odd} & \\ \hline 
 & v v \tilde{S} \tilde{S} & {\bf 0} & n \; \mbox{even} \\
\underline{{\bf m}}(\vec{q},t) = & & & \\
 & {\bf 0} & v v \tilde{S} \tilde{S}^{(s)} & n \; \mbox{odd}
\end{array}
\label{eq:Mblocks}
\end{equation}
Since this block structure is also preserved under matrix inversion the
whole set of equations is split into two parts. The first part which
consists of the block with even $n$ and $n'$ forms a closed set of equation
since also in the calculation of the memory kernels (\ref{eq:type1})
only matrix
elements with even $n$ and $n'$ occur. The second block depends on the
solution of the first set because of the structure of the memory functions
(\ref{eq:type2}) and can be shown to be 
identical to the tagged particle equations for
the self part.

\section{q-frame representation}
\label{app:q-frame}

In contrast to the case of simple liquids the density correlator ${\bf
S}(\vec{q},t)$ as defined in section \ref{sec:mmct} depends on modulus and
orientation of the vector $\vec{q}$. Therefore a direct numerical
implementation of the equations given above is not suitable but further
reformulations are necessary. The dependence on the direction of the
"outer" $\vec{q}$--vector is easily removed by choosing the $z$--axis of
the laboratory frame of reference in direction of $\vec{q}$. The choice of
the $q$--frame further offers the advantage that the matrix ${\bf
S}(\vec{q},t)$ becomes diagonal with respect to $m$ and $m'$, i.e. it holds
\begin{equation}
S_{lmn,l'm'n'}(q \hat{e}_z,t) \equiv S^m_{ln,l'n'}(q,t) \delta_{m,m'}
\equiv S_{ln,l'n'}(q,m,t) \delta_{m,m'}.
\label{eq:defqframe}
\end{equation}
To get a reformulation of the equations of motion in terms of the
$q$--frame quantities we still have to care about the "inner"
$\vec{q}$--vectors $\vec{q}_1$ and $\vec{q}_2$ appearing in the summation
of the MCT memory functions (cf. eq.(\ref{eq:memory1}).
This can be done by using the transformation
law of the tensors ${\bf S}(\vec{q},t)$ under rotations which gives a
relation between $S_{ln,l'n'}(q,m,t)$ and the molecular correlation
function for general direction of $\vec{q}$:
\begin{equation}
S_{lmn,l'm'n'}(\vec{q},t) = \sum_{m''} D^l_{m m''}(\Omega_q) D^{l'
\ast}_{m' m''}(\Omega_q) S_{ln, l'n'}(q,m'',t),
\label{eq:transform}
\end{equation}
where $\Omega_q$ denotes the polar angles of  the vector $\vec{q}$ with
respect to the laboratory frame.

Thus we get the following set of MMCT--equations in the
{\em $q$--frame representation}:
\begin{equation}
{\bf S}(q,m,z) = - \left[ z {\bf S}^{-1}(q,m) + {\bf S}^{-1}(q,m) {\bf
K}(q,m,z) {\bf S}^{-1}(q,m) \right]^{-1} \label{eq:qframe1} 
\end{equation}
\begin{equation}
K_{ln, l'n'}(q,m,z) = \sum_{\alpha \alpha'} \sum_{\mu \mu'} q^{\alpha
\mu}_{ln}(q \hat{e}_z) q^{\alpha' \mu' \ast}_{l'n'}(q \hat{e}_z) 
k^{\alpha \mu, \alpha' \mu'}_{lmn, l'mn'}(q,z) \label{eq:qframe2} 
\end{equation}
\begin{equation}
\underline{{\bf k}} = - \left[ z \underline{{\bf J}}^{-1} +
\underline{{\bf m}}(q,z) \right]^{-1} \label{eq:qframe3}
\end{equation}
with $\underline{{\bf J}}$ as given in section \ref{subsec:collective}.
The memory functions are still nondiagonal with respect to $m$ and $m'$.
But one can show that they are different from zero only in the 
following cases
\begin{equation}
\begin{array}{|@{\; \alpha=}c|@{\; \alpha'=}c||c|}
\hline 
T & T & m+\mu=m'+\mu' \\
\hline 
T & R & m+\mu=m' \\
\hline 
R & T & m    =m'+\mu' \\
\hline 
R & R & m    =m' \\
\hline
\end{array}
\end{equation}
Besides the $q$--frame representation we have used that the restricted
summation over $\vec{q}_1$ and $\vec{q}_2$, which becomes an integration in
the thermodynamic limit $V,N \to\infty$ with $\rho_0 = const$, can be reduced
to a double integral.
The general expression for the memory functions in the $q$-frame is
given by
\begin{eqnarray}
& & m^{\alpha \mu, \alpha' \mu'}_{lmn,l'm'n'}(q,t) = 
\frac{\rho_0}{(8 \pi^2)^3}
\int_0^\infty dq_1 \int_{|q-q_1|}^{q+q_1} dq_2 \frac{q_1 q_2}{q} \sum _{m_1
m_2} \sum_{l_1 l'_1 l_2 l'_2} \sum_{n_1 n'_1 n_2 n'_2} 
\; \times \nonumber \\
& & \quad \times \;
v^{\alpha \mu}_{ln, l_1 n_1, l_2 n_2}(q q_1 q_2; m m_1 m_2) \,
v^{\alpha' \mu' \ast}_{l'n', l'_1 n'_1, l'_2 n'_2}(q q_1 q_2; m' m_1 m_2) 
\; \times \nonumber \\ & & \quad \times \;
S_{l_1 n_1, l'_1 n'_1}(q_1,m_1,t) \, 
S_{l_2 n_2, l'_2 n'_2}(q_2,m_2,t) \label{eq:qframe4}
\end{eqnarray}
with the vertex functions
\begin{eqnarray}
v^{\alpha \mu}_{ln, l_1 n_1, l_2 n_2}(q q_1 q_2; m m_1 m_2) &=& 
\sum_{l_3 n_3} \left[ u^{\alpha \mu}_{ln, l_3 n_3, l_2 n_2}(q q_1 q_2; m
m_1 m_2) \, c_{l_3 n_3, l_1 n_1}(q_1, m_1) + \right. \nonumber \\
& & \left. \; + \; (-1)^m
u^{\alpha \mu}_{ln, l_3 n_3, l_1 n_1}(q q_2 q_1; m m_2 m_1) \, c_{l_3 n_3, l_2
n_2}(q_2, m_2) \right] 
\label{eq:qframe5}
\end{eqnarray}
The coefficients $u$ are given by
\begin{eqnarray}
& & u^{\alpha \mu}_{ln, l_1 n_1, l_2 n_2}(q q_1 q_2; m m_1 m_2) =
i^{l_1+l_2-l} (-1)^{m_2}
\left[ \frac{(2l_1+1) (2l_2+1)}{2l+1} \right]^{\frac{1}{2}} 
q^{\alpha \mu \ast}_{l_1 n_1}(q_1) \; \times \nonumber \\
& & \; \times \; \sum_{m''} d^{l_1}_{m'' m_1}(\theta_1)
d^{l_2}_{m-m'' m_2}(-\theta_2) {\cal C}(l_1 l_2 l; m'' \, m-m'' \, m)
\left\{ 
\begin{array}{c@{\quad}c} 
{\cal C}(l_1 l_2 l; n_1 n_2 n) & \alpha = T \\
{\cal C}(l_1 l_2 l; n_1+\mu \, n_2 \, n+\mu) & \alpha = R
\end{array} \right. .
\label{eq:qframe6}
\end{eqnarray}
%
%
%
The quantities $d^l_{m n}$ are the reduced Wigner matrices \cite{G&G}
and the angles $\theta_i$ are determined by
\begin{equation}
\cos(\theta_1) = \frac{q^2+q_1^2-q_2^2}{2 q q_1} \; , \; \sin(\theta_1) =
\sqrt{1 - \cos(\theta_1)^2}
\label{eq:angles}
\end{equation}
and the corresponding relations with $(1 \leftrightarrow 2)$. 
Further we have
\begin{equation}
q_{ln}^{\alpha \mu}(q_i)= \left\{
\begin{array}{c@{\quad}c}
\frac{1}{\sqrt{2}} q_i \sin(\theta_i) & \alpha = T, \mu =
\pm 1 \\ q_i \cos(\theta_i) & \alpha = T, \mu = 0 \\ 
& \\
\frac{1}{\sqrt{2}} \sqrt{l(l+1)-n(n+\mu)} &
\alpha = R, \mu \pm 1 \\ 
n & \alpha = R, \mu = 0 \\ \end{array} \right.
\label{eq:qcoeffs}
\end{equation}

It should be immediately obvious that a numerical solution of the equations
of motion given here poses a formidable task. Due to the large number of
summations and the occurrence of special functions the evaluation of the
memory functions will be the main computational problem.

The first step in an analysis of the equations for ${\bf S}(q,m,t)$ is the
localization of the critical temperature $T_c$ at which a bifurcation of
the long--time behavior of the solutions takes place. Therefore one studies
the nonergodicity parameters
\begin{equation}
{\bf F}(q,m) = \lim_{t \to\infty} {\bf S}(q,m,t) = - \lim_{z \to 0} z {\bf
S}(q,m,z)
\label{eq:nonerg}
\end{equation}
which obey the following equations
\begin{eqnarray}
& & {\bf F}(q,m) = \left[ {\bf S}^{-1}(q,m) + {\bf S}^{-1}(q,m) {\bf K}(q,m)
{\bf S}^{-1}(q,m) \right]^{-1} \label{eq:nonerg1} \\
& & K_{ln, l'n'}(q,m) = \sum_{\alpha \alpha'} \sum_{\mu \mu'} q^{\alpha \mu
}_{ln}(q) q^{\alpha' \mu' \ast}_{l'n'}(q) \left( \underline{m}^{-1}(q)
\right)^{\alpha \mu, \alpha' \mu'}_{lmn, l'mn'} \label{eq:nonerg2} \\ 
& & m^{\alpha \mu, \alpha' \mu'}_{lmn, l'm'n'}(q) = 
\frac{\rho_0}{(8 \pi^2)^3}
\int_0^\infty dq_1 \int_{|q-q_1|}^{q+q_1} dq_2 \sum_{m_1 m_2} \sum_{l_1
l'_1 l_2 l'_2} \sum_{n_1 n'_1 n_2 n'_2} 
\; \times \nonumber \\ & & \quad \times \; 
v^{\alpha \mu}_{ln, l_1 n_1, l_2 n_2}(q q_1 q_2; m m_1 m_2) \,
v^{\alpha' \mu' \ast}_{l'n', l'_1 n'_1, l'_2 n'_2}(q q_1 q_2; m' m_1 m_2) \;
\times \nonumber \\ & & \quad \times \; 
F_{l_1 n_1, l'_1 n'_1}(q_1,m_1,t) \, F_{l_2 n_2, l'_2 n'_2}(q_2,m_2,t) 
\label{eq:nonerg3}
\end{eqnarray}
that have to be solved self--consistently.

\end{appendix}




\begin{figure}
\caption{Critical center of mass nonergodicity parameter $F_{00,00}(q,m=0)$ as
calculated from the MD simulation (symbols) compared with the theoretical
predictions obtained in two different approximation schemes:
MMCT in the diagonal-dipole approximation
(dashed line) and MMCT in the pure dipole approximation (dotted line).}
\label{fig:fqZZZZZZ} 
\end{figure}


\begin{figure}
\caption{Diagonal critical nonergodicity parameters with $l=l'=1$
($F_{10,10}(q,m)$) as calculated from the MD simulation (symbols) compared
with the theoretical predictions of MMCT in the diagonal-dipole
approximation (dashed line) and MMCT in the pure dipole approximation
(dotted line)} 
\label{fig:fig1mct} 
\end{figure}


\begin{figure}
\caption{Diagonal critical nonergodicity parameters with $l=l'=2$
($F_{20,20}(q,m)$) as calculated from the MD simulation (symbols) compared
with the theoretical predictions of MMCT in the diagonal-dipole
approximation (dashed line) and MMCT in the pure dipole approximation
(dotted line)} 
\label{fig:fig2mct} 
\end{figure}

\begin{figure}
\caption{Off-diagonal critical nonergodicity parameters ($l \ne l'$) as
calculated from the MD simulation (symbols) compared with the theoretical
predictions of MMCT in the pure dipole approximation (dotted line)}
\label{fig:figcrossmct} 
\end{figure}


\setcounter{figure}{0}

\eject

\begin{figure}
\centerline{\psfig{figure=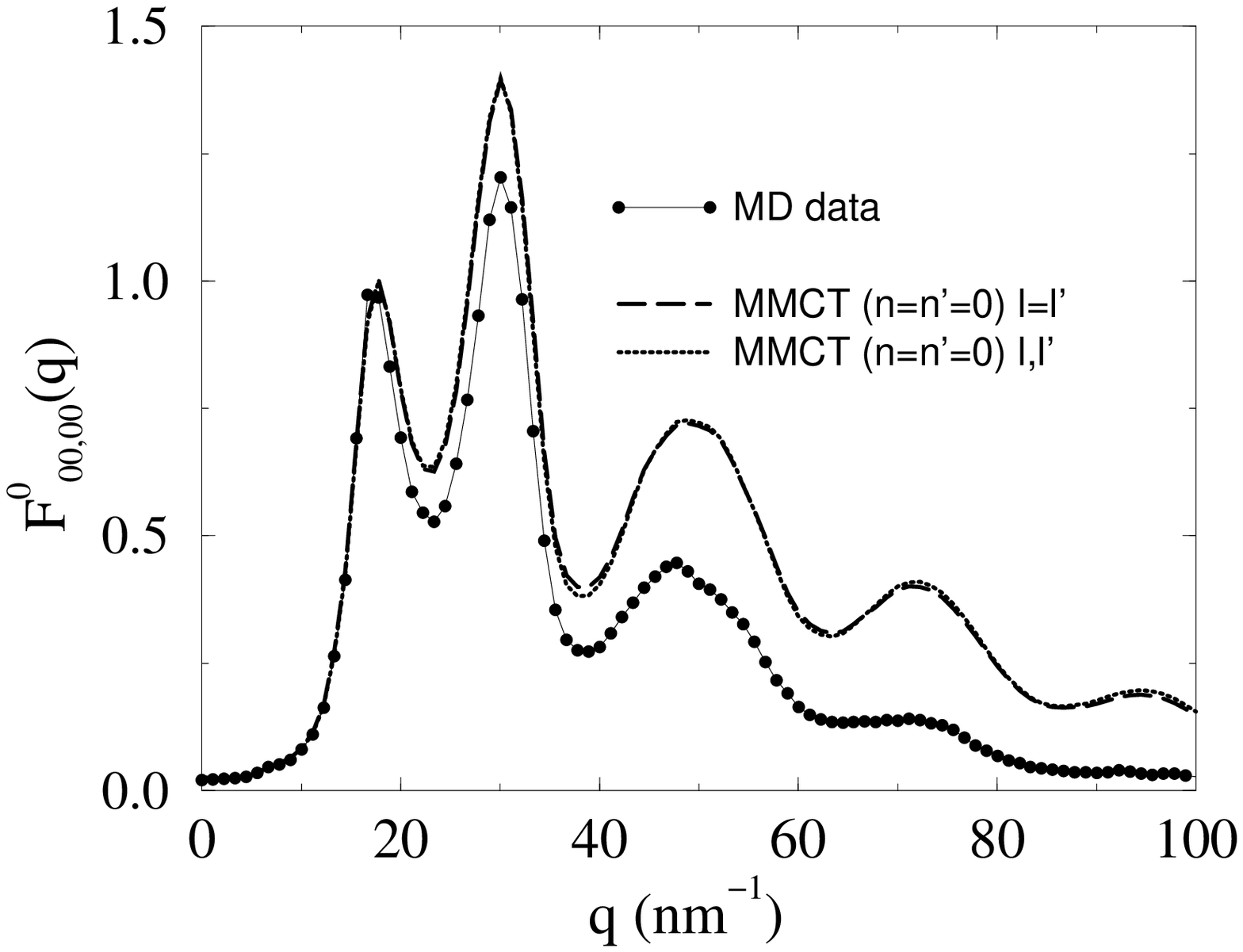,height=16cm,width=16cm,clip=,angle=0.}}
\caption{L. Fabbian et al}
\end{figure}


\begin{figure}
\centerline{\psfig{figure=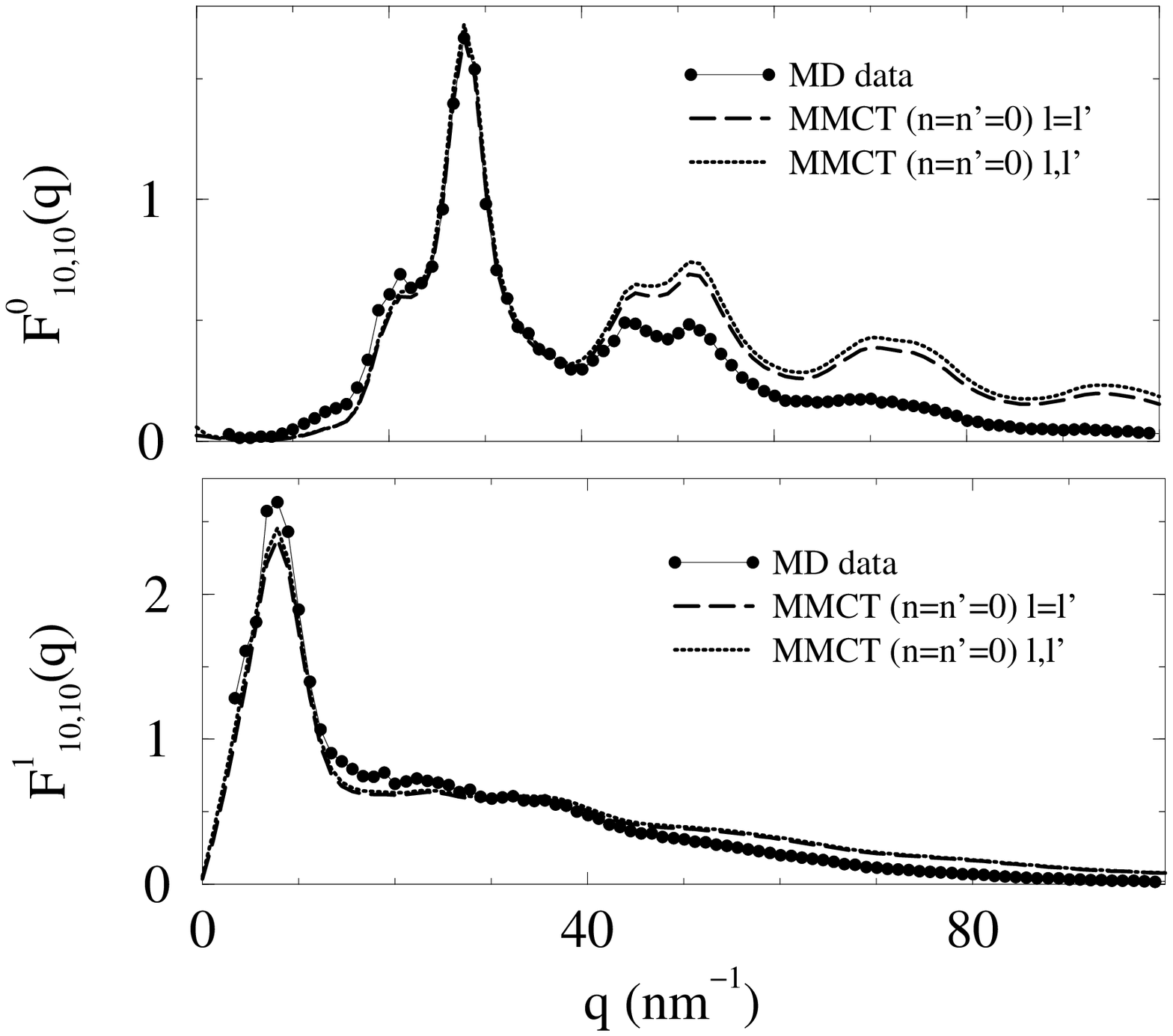,height=16cm,width=16cm,clip=,angle=0.}}
\caption{L. Fabbian et al}
\end{figure}

\begin{figure}
\centerline{\psfig{figure=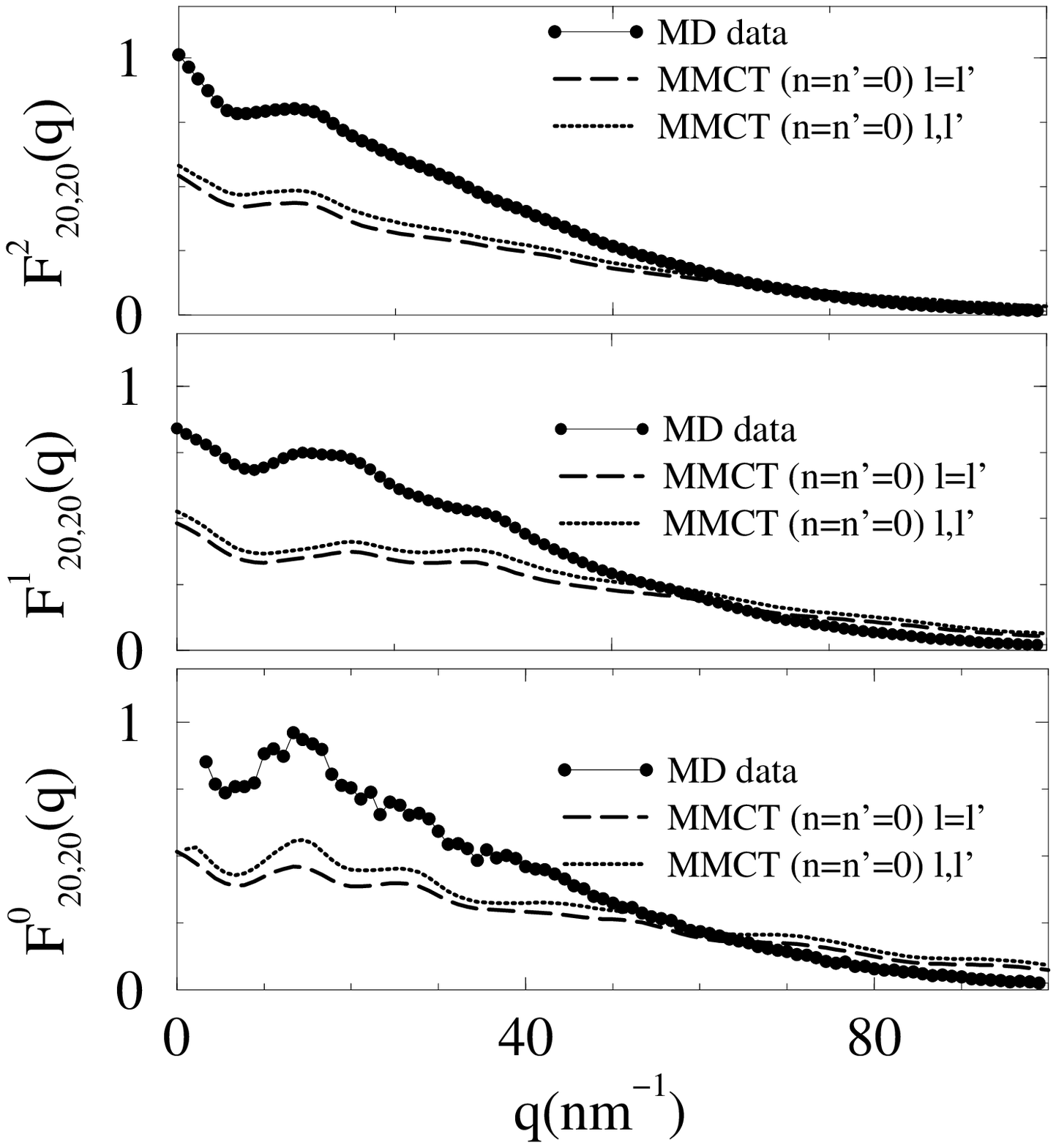,height=16cm,width=16cm,clip=,angle=0.}}
\caption{L. Fabbian et al.}
\end{figure}

\begin{figure}
\centerline{\psfig{figure=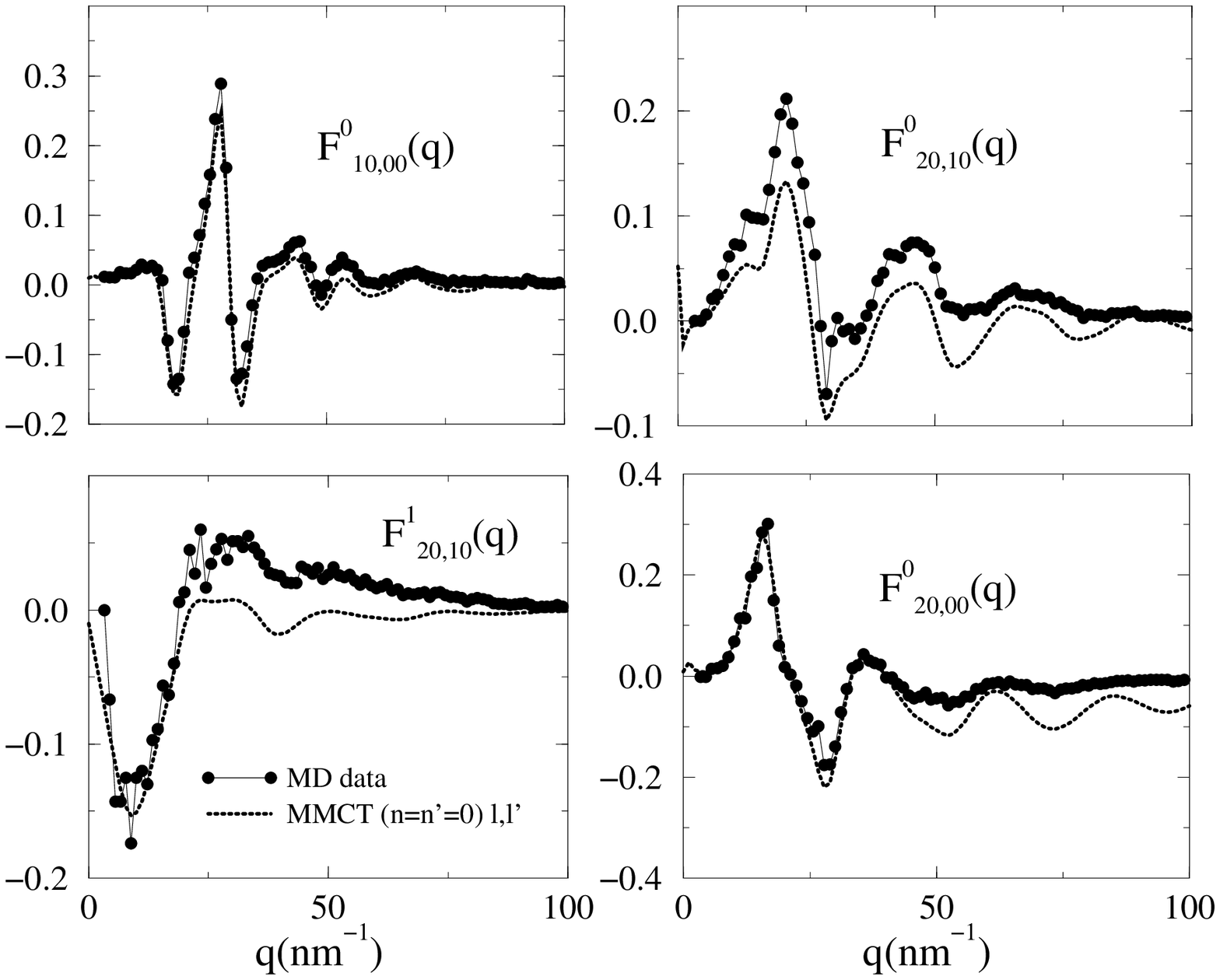,height=16cm,width=16cm,clip=,angle=0.}} 
\caption{L. Fabbian et al}
\end{figure}


\end{document}